\documentclass[12pt,a4paper]{article}
\usepackage{amsmath}
\usepackage[dvipdfmx]{graphicx}
\begin{document}

\title{\huge{A study of silicon sensor for ILD ECAL}}
\author{\Large{Tatsuhiko Tomita\thanks{tomita@epp.phys.kyushu-u.ac.jp}~\thanks{Kyushu University}} \\ Shion Chen\thanks{The University of Tokyo}, ~Daniel Jeans$^{\ddagger}$, ~Yoshio Kamiya\thanks{International Center for Elementary Particle Physics, The University of Tokyo}, \\
Kiyotomo Kawagoe$^{\dagger}$, ~Sachio Komamiya$^{\ddagger}$, ~Chihiro Kozakai$^{\ddagger}$, \\
Yohei Miyazaki$^{\dagger}$, ~Taikan Suehara$^{\dagger}$, ~Yuji Sudo$^{\dagger}$,\\ Hiraku Ueno$^{\dagger}$, ~Tamaki Yoshioka$^{\dagger}$}
\date{}
\maketitle

\begin{abstract}
\thispagestyle{empty}
\noindent The International Large Detector (ILD) is a proposed detector for the
International Linear Collider (ILC) \cite{TDR,TDRD}. It has been designed to achieve an excellent jet
energy resolution by using Particle Flow Algorithms (PFA) \cite{PFA}, which rely on the ability to separate nearby particles within jets. PFA requires calorimeters with high granularity. The ILD Electromagnetic Calorimeter (ECAL) is a sampling calorimeter with thirty tungsten absorber layers. The total thickness of this ECAL is about 24 X$_0$, and it has between 10 and 100 million channels to make high granularity. Silicon sensors are a candidate technology for the sensitive layers of this ECAL. Present prototypes of these sensors have 256 5.5$\times$5.5 mm$^2$ pixels in an area of 9$\times$9cm$^2$.We have measured various properties of these prototype sensors: the leakage current, capacitance, and full depletion voltage. We have also examined the response to an infrared laser to understand the sensor's response at its edge and between pixel readout pads, as well the effect of different guard ring designs.
\\
\noindent In this paper, we show results from these measurements and discuss future works.
\end{abstract}

\footnote[0]{Talk presented at the International Workshop on Future Linear Colliders (LCWS13), Tokyo, Japan, 11-15 November 2013. }

\newpage

\section{Introduction}

The International Linear Collider (ILC) is a future high energy lepton collider \cite{TDR}. It can make precision measurements of the Higgs boson, top and electroweak physics. ILC has a capability of exploring new physics as well. To realize these measurements, we should prepare a detector with excellent precision \cite{TDRD}. 

In the case of ILC, many events will include jets in their final state. For precise measurements, these jets should be reconstructed as well as possible. We plan to use Particle Flow Algorithm (PFA) \cite{PFA} to achieve Jet Energy Resolution (JER) 3-4\%.
To make full use of PFA, we should separate each shower made from individual particles in jets. Thus, we need high granular calorimeter especially in ECAL. 

In this paper, we show basic properties of silicon sensors which should be measured for quality control, and we also report the response to an infrared laser.

\section{Overview of silicon sensors for ILD ECAL}
The silicon sensor chips are produced by Hamamatsu (HPK). The sensors are a kind of silicon PIN diode, and they have 256 5.5$\times$5.5mm$^2$ pixels in an area of 9$\times$9cm$^2$ (Figure~1). The thickness of sensors is 320 $\mu$m. The guard ring can collect surface current of the chips, but it also limit sensitive region of chips. Furthermore, from a former test beam result, 1 guard ring structure makes fake signals along it when particles come into near guard ring. In this way, guard ring has advantage and disadvantage, thus we have to decide which design should be used for ILD ECAL.

\begin{figure}[hhh]
\begin{minipage}{0.5\hsize}
\begin{center}
\includegraphics[width=80mm,height=60mm]{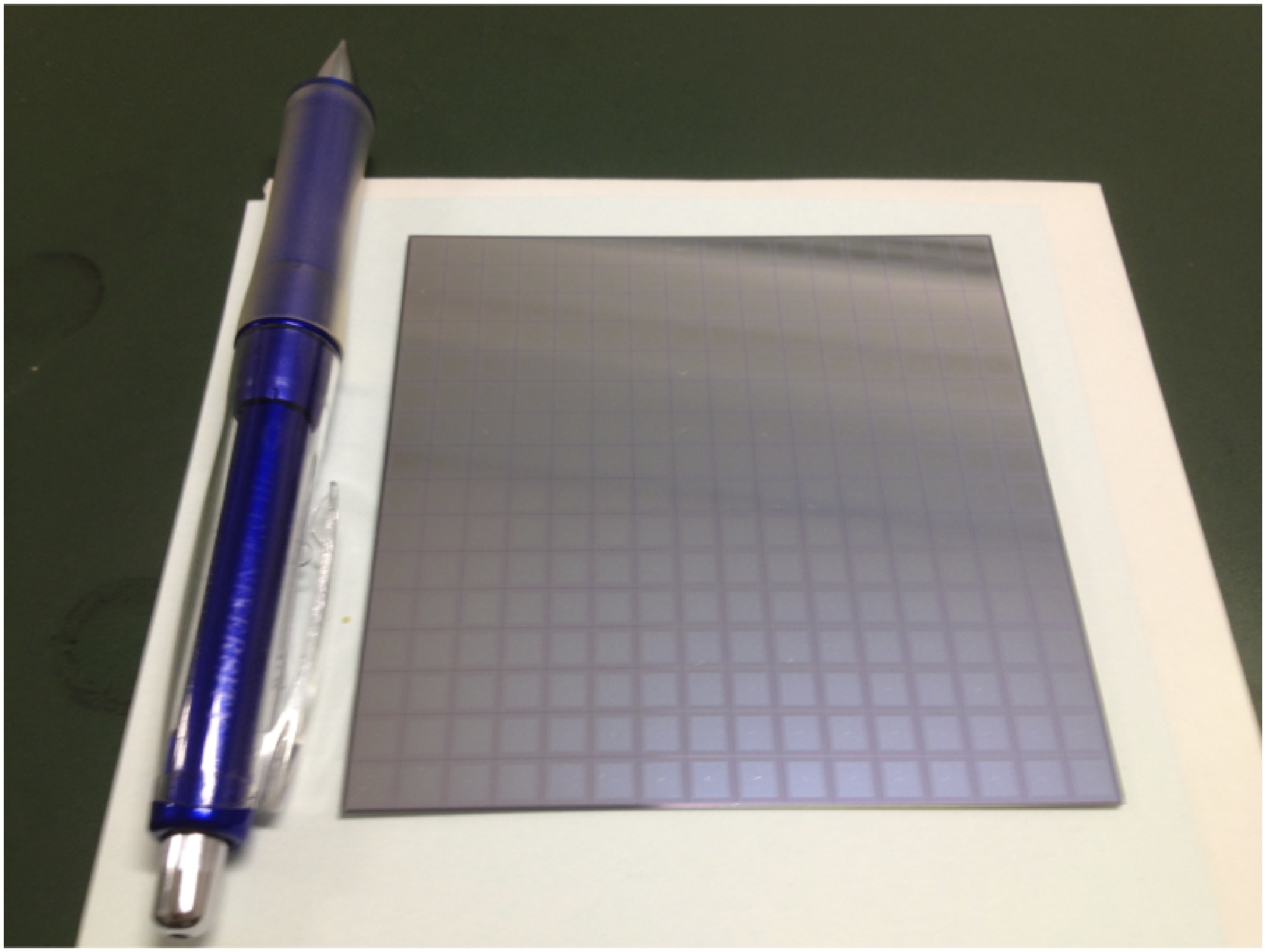}
\end{center}
\caption{Silicon sensor: each chip has 256 channels.}
\label{fig:sensor}
\end{minipage}%
\hspace{2pt}
\begin{minipage}{0.5\hsize}
\begin{center}
\includegraphics[width=80mm,height=70mm,bb= 0 0 1024 768]{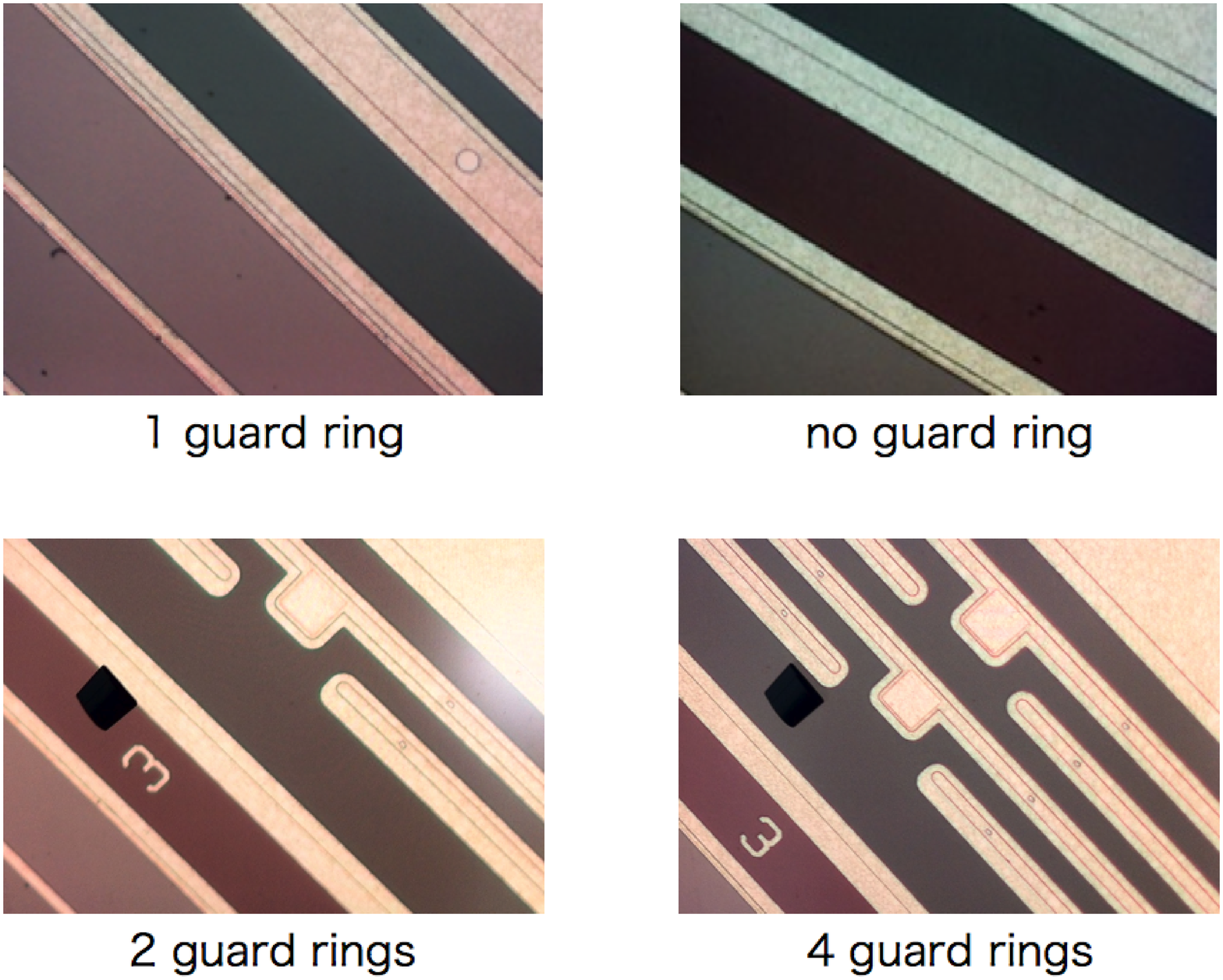}
\end{center}
\caption{Guard ring structures:  the guard ring width is 80 $\mu$m (1 guard ring), or 20 $\mu$m (2, 4 guard rings).}
\label{fig:gr}
\end{minipage}
\end{figure}
\clearpage

There are four types of guard ring structure. Currently, we have only 1 guard ring structure in this size. The other structures are provided by smaller chips 3$\times$3 pixels (4 guard rings), or 4$\times$4 pixels (no guard ring, 2 guard rings). We measured only 1 guard ring type chip, but in the near future we will measure the other chip and compare the properties of each structure.

\section{Basic properties: leakage current and capacitance}
We made two setups for the measurements of basic properties. One of the them is for the measurement of leakage current, the other is for capacitance measurement (Figure~3).

\begin{figure}[hhh]
\begin{center}
\includegraphics[width=120mm,height=70mm,bb= 0 0 1024 768]{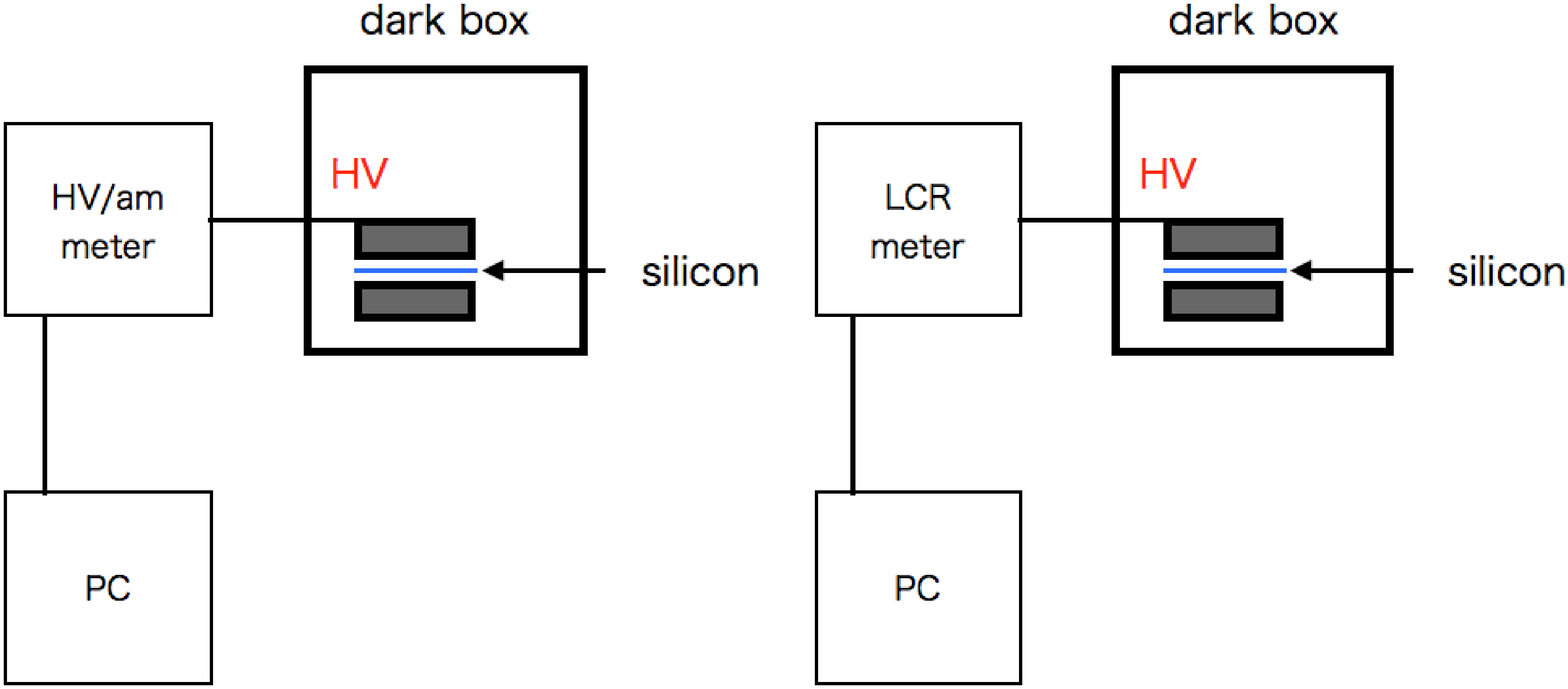}
\end{center}
\caption{Setup: the left one is for leakage current, the other one is for capacitance.}
\label{fig:setup}
\end{figure}
\begin{figure}[hhh]
\begin{minipage}{0.5\hsize}
\begin{center}
\includegraphics[width=80mm,height=65mm,bb= 0 0 1024 768]{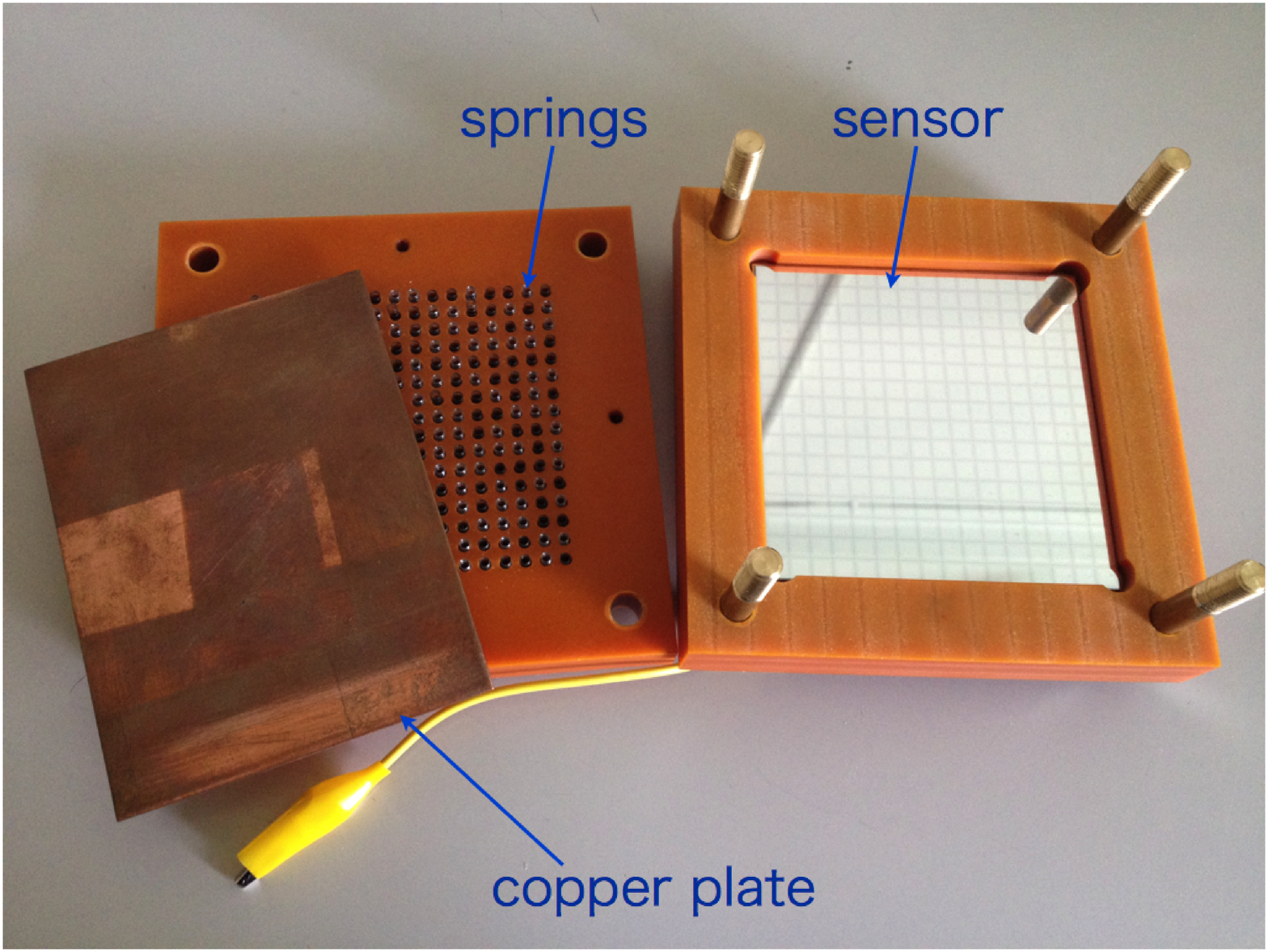}
\end{center}
\caption{Setup box: copper plate combine 256 channels to 1 channel.}
\label{fig:box}
\end{minipage}%
\hspace{2pt}
\begin{minipage}{0.5\hsize}
\begin{center}
\includegraphics[width=80mm,height=65mm,bb= 0 0 1024 768]{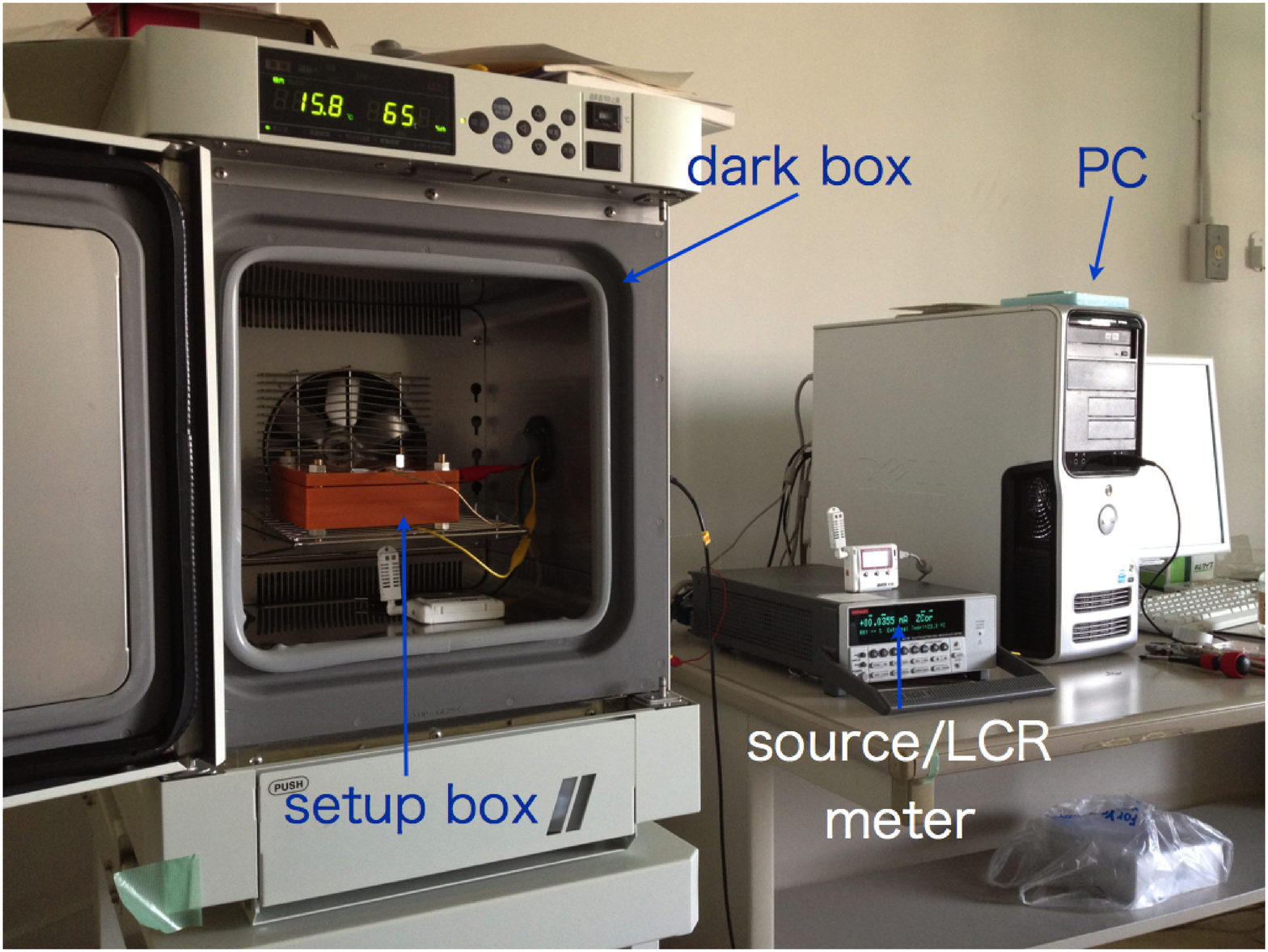}
\end{center}
\caption{Setup for the measurements: source/LCR meter is controlled from a PC.}
\label{fig:darkbox}
\end{minipage}
\end{figure}

\clearpage

We can control temperature and humidity of dark box, thus we are able to assume these two dependence are negligible in both measurements. In our measurements, we set temperature at 27.5 $^{\circ}$C (or 20.0 $^{\circ}$C for capacitance measurement) and humidity at 50\%. During both measurements, we used copper plate to combine signals from all sensor cells channel at one readout channel (see Figure~4). Between sensor and read out circuit are connected by using springs. The operation and data taking were controlled from a PC (see Figure~5).

For leakage current measurement, we took data from 0 V to 800 V in 10 V step. Each step has 1 second, and if the chip reaches to breakdown voltage we stop the measurement to prevent the chip from getting damaged. As shown in Figure~6, the leakage current is about 100 nA with applied bias voltage of 100-150 V, and its voltage dependence is very small in this region. We repeated this procedure 10-20 times on each chip, and found that the breakdown voltage becomes higher than the previous scan at first 3-4 scans. This phenomenon needs to be understood. However, we can conclude the leakage current is quite small and stable at the operation voltage of 100-150 V. If a chip has bad channel of high leakage current, we expect that the chip reaches breakdown on much smaller voltage. Therefore, this measurement is also important to test the chips before assembly, especially in mass production phase.

\begin{figure}[hhh]
\begin{minipage}{0.5\hsize}
\begin{center}
\includegraphics[width=80mm,height=80mm,bb= 0 0 1024 768]{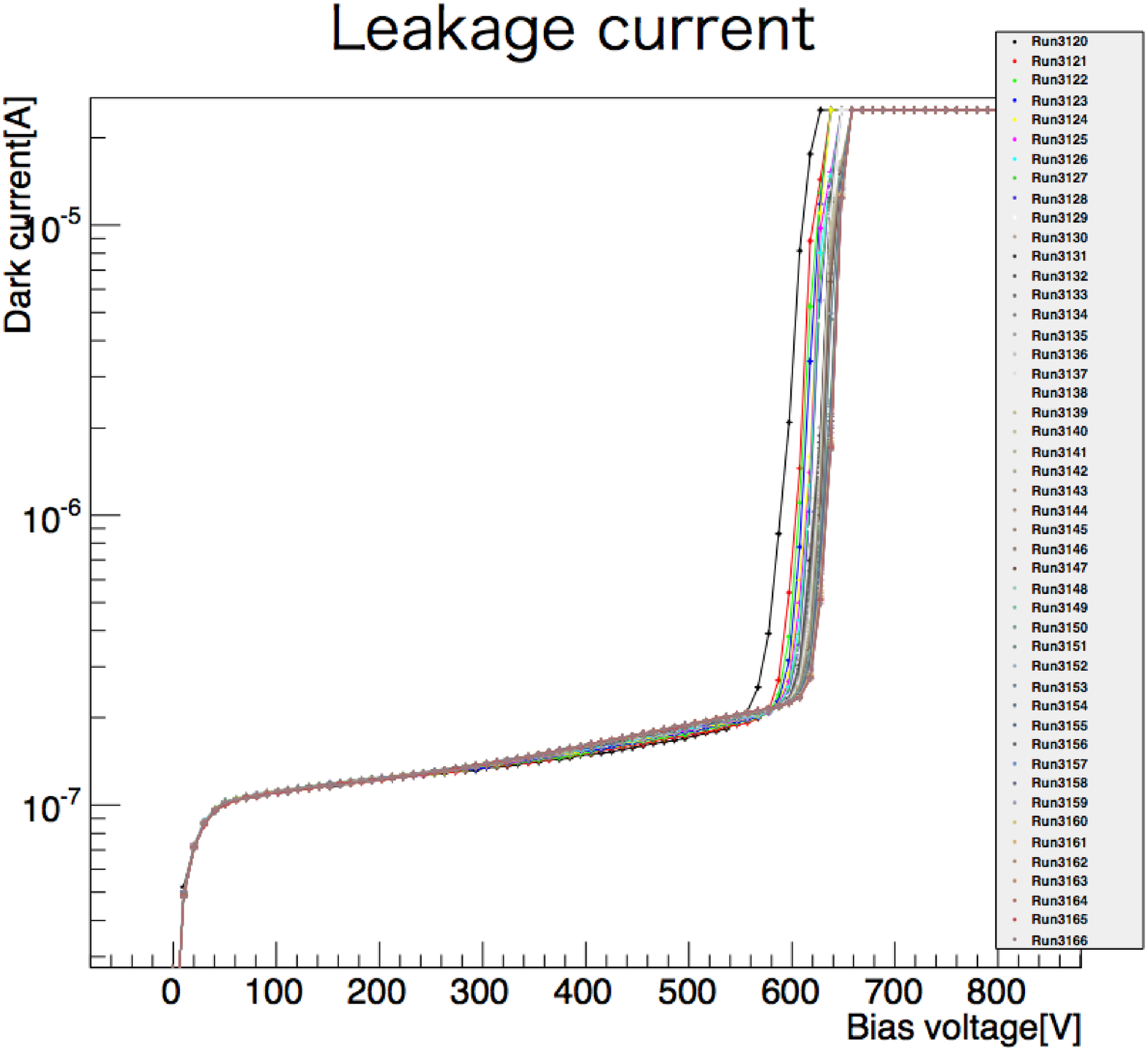}
\end{center}
\caption{Leakage current of the chip (0-800 V) under the situation of 27.5 $^{\circ}$C. Horizontal axis shows applied voltage, and vertical axis shows the total leakage current of silicon sensor. 47 scans of the same chip are overlaid.}
\label{fig:leakage}
\end{minipage}%
\hspace{2pt}
\begin{minipage}{0.5\hsize}
\begin{center}
\includegraphics[width=80mm,height=80mm,bb= 0 0 1024 768]{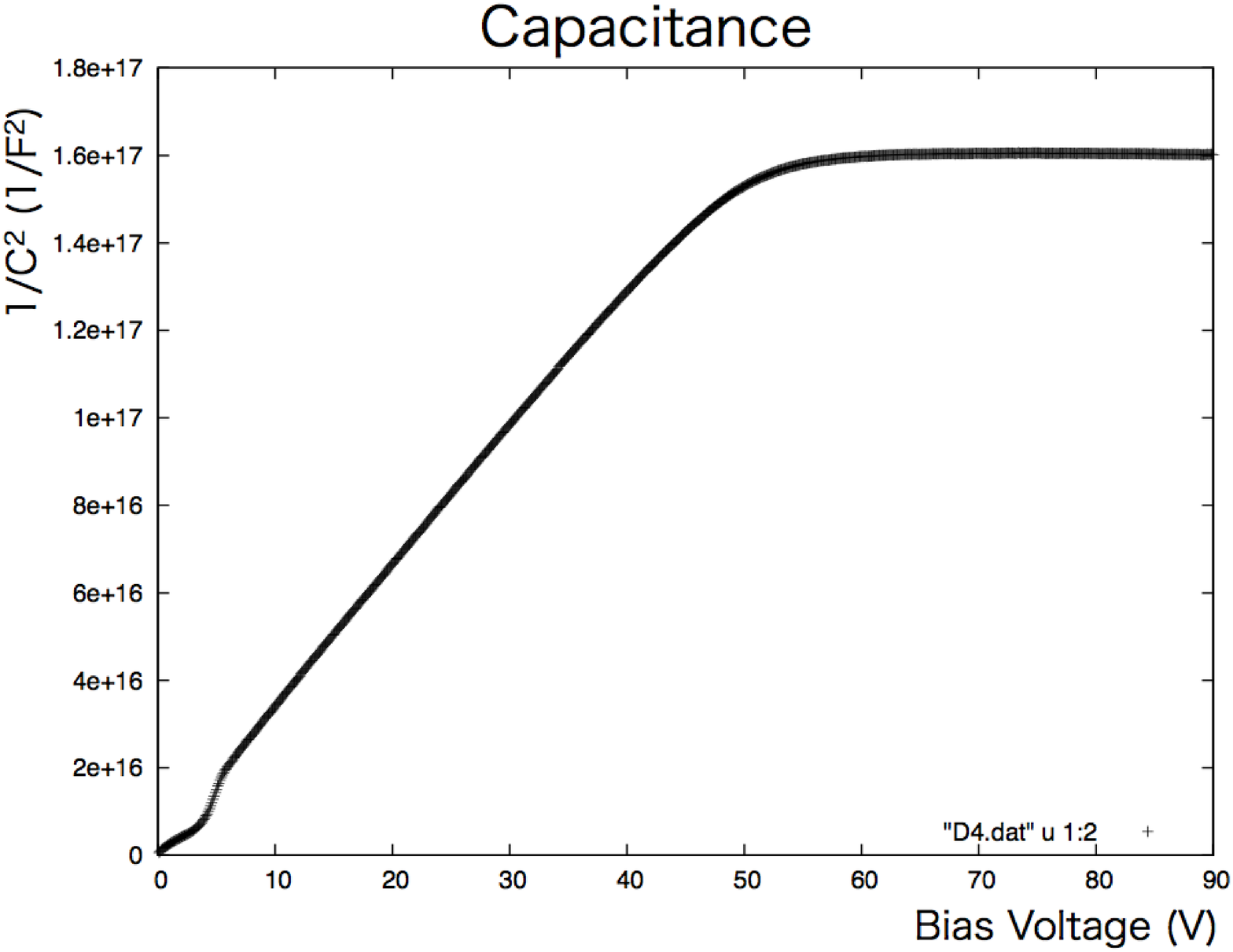}
\end{center}
\caption{Capacitance curve: the horizontal axis shows applied voltage, and vertical axis shows 1/capacitance$^2$. This measurement did in the situation of 20 $^{\circ}$C and 50 \% of humidity.}
\label{fig:capacitance}
\end{minipage}
\end{figure}

\newpage

For capacitance measurement, we applied bias voltage from 0.1 V to 130 V by step of 0.1 V. Figure~7 shows good linearity in 5 V to 45 V region. From 0 V to 5 V, the capacitance curve is slightly winding because of its Metal-Oxide Semiconductor structure \cite{MOS}. Shown in Figure~7, the capacitance saturates at around 60 V, which is consistent of the full depletion voltage of 65 V in the specification. We also estimated that the chip thickness is 318$\mu$m. This value is also consistent with the specification.

\section{Response to the laser}
To understand the difference between different guard ring structures, we prepared a laser system. We use an infrared laser (class-3B) whose wavelength is 1064 nm. Photons of this wavelength can pass through the surface of silicon with high transmission to produce a pair of electron-hole in the active area. Since infrared photons cannot go through electrode, we shoot the laser into the gap (between a pixel and a pixel, or a pixel and a guard ring). 

At first, we measured the response from a specific pixel to check our system. We used an oscilloscope to see waveform, and we also check the fluctuation of peak value by using the waveform. The waveform taken at 100 V bias voltage is shown in Figure~8. The decay tail was caused by the time constant of the pre-amplifier. We obtained the peak fluctuation value of ~3\%, shown in Figure~9, which corresponds to the stability of our measurement.

\begin{figure}[hhh]
\begin{minipage}{0.5\hsize}
\begin{center}
\includegraphics[width=80mm,height=80mm,bb= 0 0 1024 768]{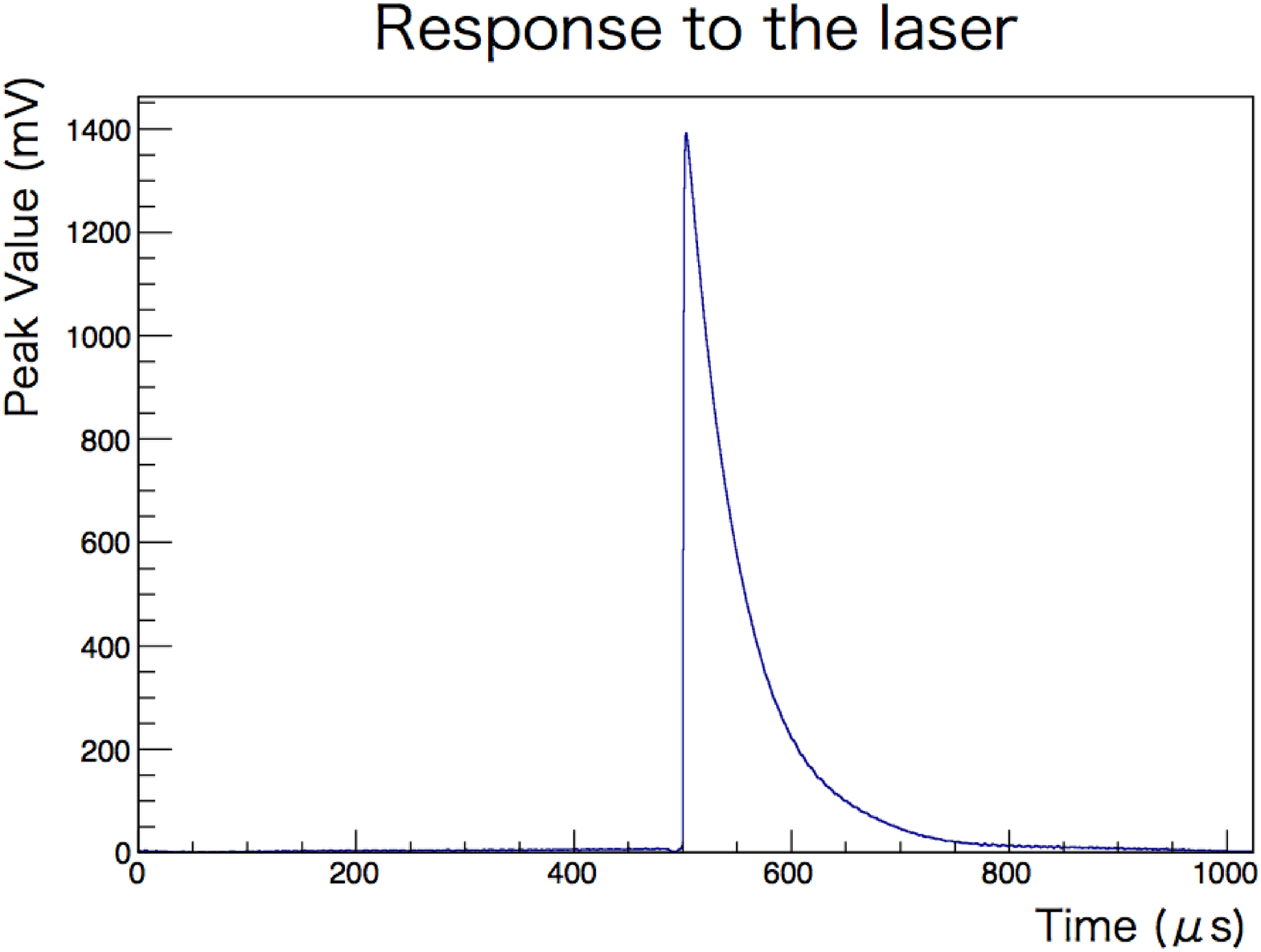}
\end{center}
\caption{Wave form: the horizontal axis shows time ($\mu$s), vertical axis shows output value (mV) time constant of pre-amplifier is about 150~$\mu$s. This measurement ran under applied voltage of 100 V.}
\label{fig:wave}
\end{minipage}%
\hspace{2pt}
\begin{minipage}{0.5\hsize}
\begin{center}
\includegraphics[width=85mm,height=85mm,bb= 0 0 1024 768]{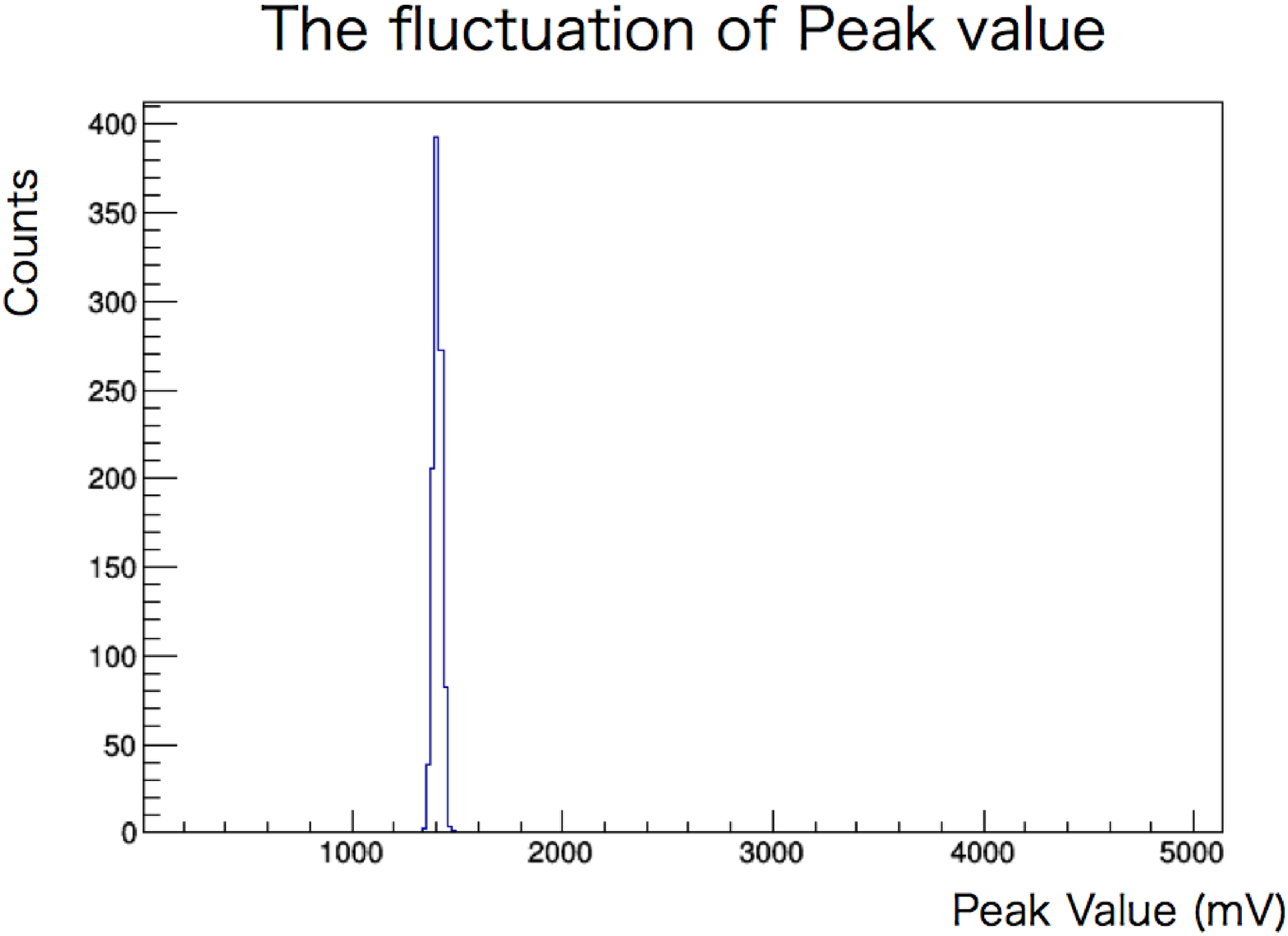}
\end{center}
\caption{Peak fluctuation: the horizontal axis shows peak value of response, and vertical axis shows the number of counts. The measurement did in 100 V applied voltage.}
\label{fig:fluc}
\end{minipage}
\end{figure}

\clearpage

\begin{figure}[hhh]
\begin{center}
\includegraphics[width=120mm,height=90mm,bb= 0 0 1024 768]{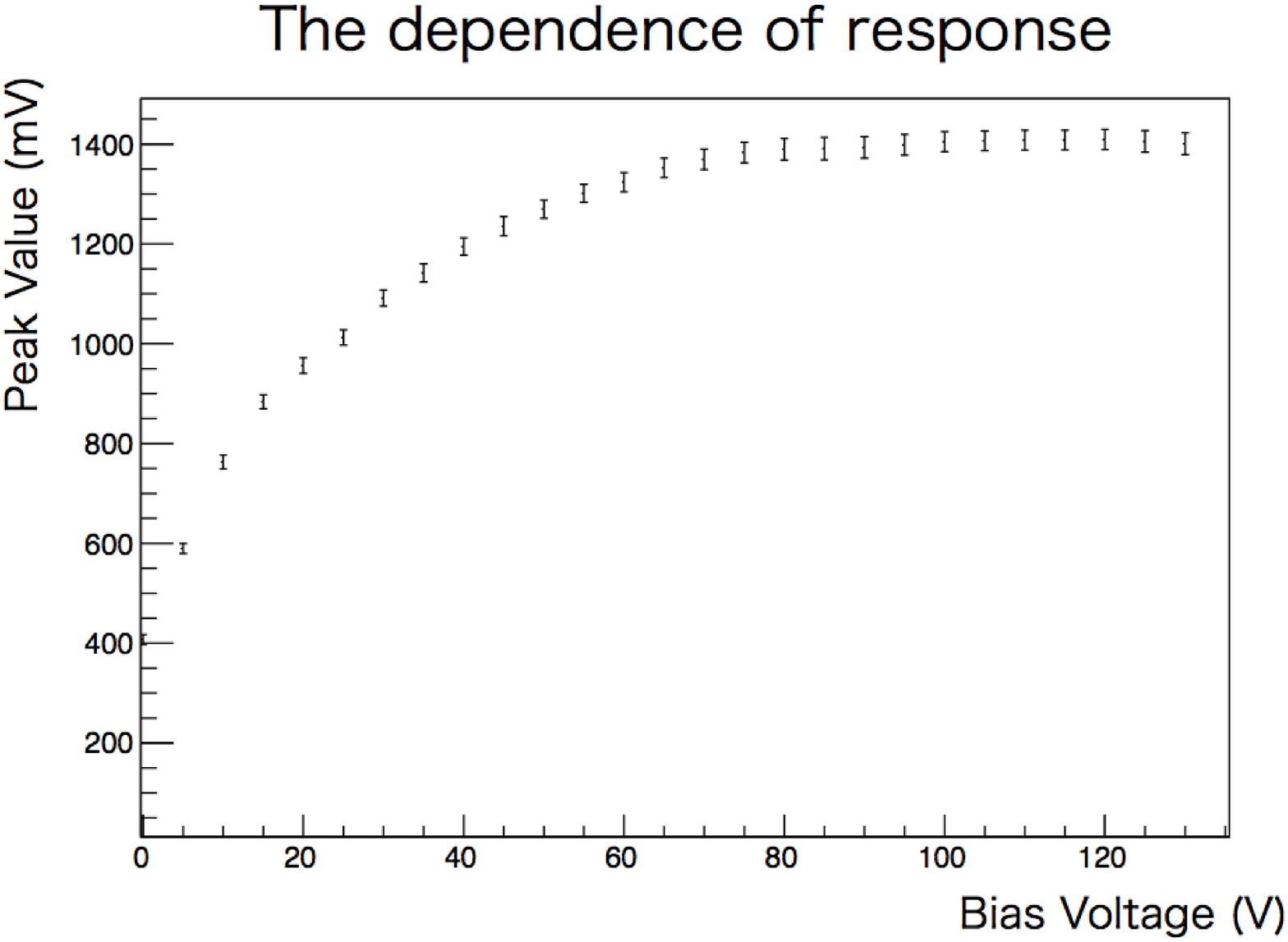}
\end{center}
\caption{The voltage dependence of peak value: The horizontal axis shows applied voltage, and vertical axis shows peak value. The error bar comes from the fluctuation of peak value (shown in Figure~9).}
\label{fig:voldep}
\end{figure}

We also investigated the response for various bias voltages. The result is shown in Figure~10. In this figure, the response saturates at 80 V, which indicates that the chip reached full depletion. Since this saturation is not consistent to the full depletion voltage of 65 V (obtained from previous measurements), we are now investigating the reason of the difference.

\section{Summary}
We established a silicon sensor test system. We can measure basic properties of silicon sensors such as leakage current and capacitance. Our first result from these measurements meet the specification. We also investigated the response to an infrared laser. We have a small disagreement for the full depletion voltage between the capacitance measurement and the laser measurement. Now we are trying to understand this issue. Our final goal is to make a decision on the chip design including guard ring structures, thus we will compare these properties on prototypes of various structures in the near future.

\end{document}